\documentclass{osa-article}

\journal{oe}

\articletype{Research Article}

\usepackage{lineno}

\begin{document}

\title{Optical Frequency Combs in Aqueous and Air Environments at Visible to Near-IR Wavelengths}

\author{Gwangho Choi,\authormark{1} Adley Gin,\authormark{1} and Judith Su\authormark{1,2,*}}

\address{\authormark{1}College of Optical Sciences, University of Arizona, Tucson, Arizona 85721, USA\\
\authormark{2}Department of Biomedical Engineering, University of Arizona, Tucson, Arizona 85721, USA}

\email{\authormark{*}judy@optics.arizona.edu} 



\begin{abstract}
The ability to detect and identify molecules at high sensitivity without the use of labels or capture agents is important for medical diagnostics, threat identification, environmental monitoring, and basic science. Microtoroid optical resonators, when combined with noise reduction techniques, have been shown capable of label-free single molecule detection, however, they still require a capture agent and prior knowledge of the target molecule. Optical frequency combs can potentially provide high precision spectroscopic information on molecules within the evanescent field of the microresonator; however, this has not yet been demonstrated in air or aqueous biological sensing. For aqueous solutions in particular, impediments include coupling and thermal instabilities, reduced Q factor, and changes to the mode spectrum. Here we overcome a key challenge toward single-molecule spectroscopy using optical microresonators: the generation of a frequency comb at visible to near-IR wavelengths when immersed in either air or aqueous solution. The required dispersion is achieved via intermodal coupling, which we show is attainable using larger microtoroids, but with the same shape and material that has previously been shown ideal for ultra-high sensitivity biosensing. We believe that the continuous evolution of this platform will allow us in the future to simultaneously detect and identify single molecules in both gas and liquid at any wavelength without the use of labels.
\end{abstract}

\section{Introduction}
Optical microtoroid resonators are attractive biochemical sensors due to their ultra-high quality (Q) factors and small mode volumes~\cite{kippenberg_kerr-nonlinearity_2004,herr_temporal_2014,kippenberg_dissipative_2018,gaeta_photonic-chip-based_2019,ozgur_ultrasensitive_2019,li_dark_2019,chen_simulating_2019,su_label-free_2015}. In addition, microtoroids have a larger capture area compared to nanoscale sensors such as nanorods and nanowires thus making detection events more likely~\cite{suebka_how_2021}. We have previously demonstrated that single molecules can be detected without the use of labels using a microtoroid optical resonator in combination with noise reduction techniques~\cite{su_label-free_2016,su_label-free_2015-1,hao_noise-induced_2020}. To achieve specific detection, the surface of the resonator needs to be functionalized for the target molecule of interest. A non-functionalized resonator can also detect binding events; however, in this case not only does the target molecule need to be known in advance, but the solution must either be pure or the molecules of interest must have very different binding characteristics. 

In many cases, however, target analytes are unknown and need to be identified. Since optical frequency combs can be used to identify molecular species~\cite{coddington_dual-comb_2016,yu_microresonator-based_2017,dutt_-chip_2018}, generating a frequency comb with a microtoroid may enable detection and molecular identification on the same device without having to functionalize the surface of the sensor. Surface functionalization adds both complexity and cost to the experiment and reduces the Q-factor which reduces sensitivity. A toroidal geometry is desired for biochemical sensing over other high-Q geometries such as microdisks that have been used to generate combs as it has a large ($\sim$ 100 nm) evanescent field sensing region that is needed for biochemical sensing.

Despite the potential advantages of using frequency combs for biochemical sensing, frequency comb generation in aqueous solution has not previously been demonstrated. A key challenge is that resonator dispersion is altered significantly when an aqueous solution is injected over the resonator, making it difficult to realize the required anomalous mode dispersion. To the best of our knowledge, conventional dispersion engineering techniques have not addressed this issue~\cite{riesen_dispersion_2016,riesen_dispersion_2016-1}. Here, we generated an optical frequency comb in water and air at visible to near IR wavelengths on a microtoroid optical resonator overcoming a limit of conventional dispersion engineering. This can be achieved via an avoided mode crossing (AMX), which is an interplay between different transverse optical modes in a resonator.

\section{Dispersion Engineering and Avoided Mode Crossings}

Typically, dispersion engineering is needed to generate microresonator based frequency combs~\cite{fujii_dispersion_2020}. Total cavity dispersion is a function of material, waveguide geometry, and optical mode distribution. Material dispersion can be engineered by either replacing a material or doping it. For biosensing experiments which are performed in aqueous solutions, it is often desired to use visible or near visible wavelengths as the absorption of light in water at those wavelengths is minimized. At these wavelengths, the group velocity dispersion of a typical material is strongly normal, so the overall cavity dispersion is also normal. However, waveguide dispersion can be engineered to compensate the material dispersion~\cite{pfeiffer_photonic_2018}, resulting in  overall anomalous dispersion even at visible wavelengths. Since different optical modes have their own mode profile and effective refractive index ($n_{\text{eff}}$) the overall dispersion is also a function of optical mode~\cite{herr_mode_2014,yang_four-wave_2016,zhao_visible_2020}. 

Numerous forms of microresonator dispersion engineering have been demonstrated in the near-visible regime. In a microbubble resonator, a very thin waveguide structure was fabricated that confined the light and overcompensated the normal dispersion of the silica for the fundamental TM mode at $780~\text{nm}$~\cite{yang_four-wave_2016}. A silica microdisk with a large wedge angle was shown to achieve an anomalous dispersion at $780~\text{nm}$ by controlling the angle of the wedge (a form of waveguide dispersion)~\cite{ma_visible_2019}. Anomalous dispersion was also generated in a silicon-nitride integrated ring resonator by using a high radial order mode (modal dispersion)~\cite{zhao_visible_2020,domeneguetti_parametric_2021}. These demonstrations are all based on a fixed cladding material; however, varying the surrounding material can alter the overall dispersion.

While tight confinement of light can overcompensate material dispersion in the visible and NIR wavelength regime, a high refractive index contrast between the waveguide material (i.e., silica) and the cladding (i.e., air) is needed. If the surrounding air is replaced by a liquid (n > 1.33), the light loses strong confinement, and the normal dispersion of the material cannot be compensated (see Supplementary Information Section 1 for more details). Several ways to overcome this issue include adding a high index material coating on the resonator surface~\cite{jin_dispersion_2017,wang_whispering-gallery_2018,riemensberger_dispersion_2012}, engineering the modal dispersion of a cavity~\cite{zhao_visible_2020}, or replacing the waveguiding material by a higher index material~\cite{riesen_material_2015,riesen_dispersion_2016,riesen_dispersion_2016-1}.

If, however, it is desired to preserve the material and structure of a resonator, a different approach is needed. A distinct property of a resonator where different optical modes have their own free spectral range (FSR) is that they can interact and couple with one another resulting in deviations from the original, unperturbed  FSR~\cite{carmon_static_2008,savchenkov_kerr_2012,herr_mode_2014,xue_thermal_2016}. This localized dispersion perturbation can lead to a local anomalous dispersion which can meet a phase-matching condition for the four wave mixing (FWM) process. At visible or near visible wavelengths, optical frequency combs have been  generated via AMXs in a crystalline WGM resonator~\cite{savchenkov_kerr_2012}, a microring resonator~\cite{karpov_photonic_2018,nazemosadat_switching_2021,liu_investigation_2014}, dual microring resonators~\cite{xue_normal-dispersion_2015,miller_tunable_2015}, and a wedge disk resonator~\cite{lee_chemically_2012,lee_towards_2017,wang_dirac_2020}. Chip-based ring resonators can be designed to introduce AMXs at desired locations using a thermal heater~\cite{xue_mode-locked_2015} or by adding an another resonator nearby~\cite{kim_dispersion_2017,xue_normal-dispersion_2015,miller_tunable_2015}. Although toroidal resonators have been one of main platforms in biosensing experiments, there has been no demonstration of optical frequency comb generation under a sensing environment where the toroid is immersed in liquid.

\section{Device Fabrication \& Dispersion Measurement}

\begin{figure*}[ht!]
\centering
\includegraphics[width=\linewidth]{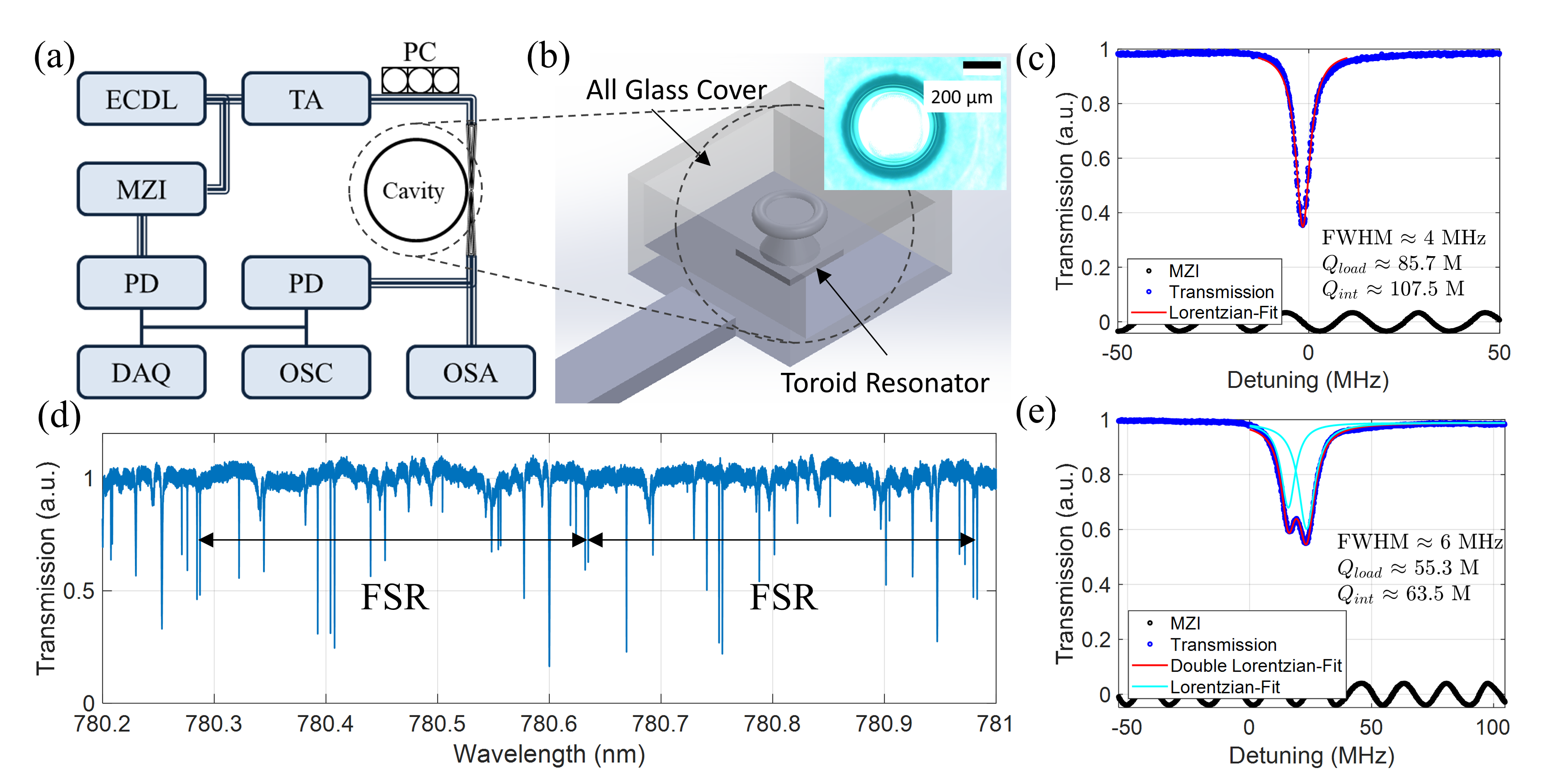}
\caption{(a) Experimental Setup. An external cavity diode laser (ECDL) is amplified by a tapered amplifier (TA) and pumped into a cavity. A polarization controller (PC) is used to excite either the TE or TM mode family. Laser wavelength scanning is calibrated using a Mach-Zehnder interferometer (MZI). The calibration and transmission data are received by photodetectors (PDs) and monitored using a data acquisition (DAQ) system or an oscilloscope (OSC). The spectrum was simultaneously measured using an optical spectrum analyzer (OSA). (b) Schematic of the sample chamber. Inset: microscope image of the microcavity. (c) Q-factor measurement in air for a microtoroid. The frequency axis is calibrated by a MZI of $FSR_{MZI}\approx17.5~\text{MHz}$. The loaded Q-factor of the fundamental mode was $\sim 1\times 10^8$. (d) Representative example of the calibrated spectrum scan. (e) Linewidth measurement for the same toroid in water. Typically, the quality factor drops by around a factor of 2. For simplicity, the linewidth of the left peak is used for the Q-factor estimation.}
\label{fig:Overview}
\end{figure*}

Currently, microtoroids with a major diameter of  $\sim100~\mu\text{m}$ are used for biosensing applications \cite{su_label-free_2016} . This diameter, however, yields a large FSR ($\sim700~\text{GHz}$ in the NIR) resulting in only tens of mode numbers in the scanning wavelength range ($\sim8~\text{THz}$) of our system. To efficiently introduce and characterize AMXs, it is highly desirable to have many modes. This can be done by fabricating a larger diameter toroid~\cite{delhaye_frequency_2009}. To fabricate larger toroids [Fig.~\ref{fig:Overview}(b)], we chose a thicker silica layer in order to avoid  stress induced defects ("buckling") caused by different thermal expansion coefficients between the silica layer and the substrate as the cavity diameter increases~\cite{chen_thermal_2013}. Buckling can significantly degrade the fabrication quality but can be controlled by proper choice of the device layer thickness and the amount of undercut and is therefore not a limiting factor for the optical performance of the device if for example, we wish to change the FSR of the resonator~\cite{lee_chemically_2012}. 

To fabricate a large ($\sim$~500-$\mu\text{m}$) diameter microtoroid, a $\sim$~700-$\mu\text{m}$ diameter disk resonator is first fabricated using conventional lithography and etching techniques~\cite{armani_ultra-high-q_2003,ma_kerr_2017,zhang_spectral_2020}. A thermally grown $6~\mu\text{m}$ thick silicon dioxide layer on top of the silicon substrate (WaferPro) was used. The amount of undercut for the microdisk needs to be large enough to isolate an optical mode from the silicon pillar and small enough to prevent buckling. By defocusing our laser beam and increasing the power of the CO$_2$ laser, we can reflow $\sim$~500-$\mu\text{m}$-diameter or larger toroids [Fig.~\ref{fig:Overview}(b)]. An ultra-high Q factor (Q > $10^8$) is routinely measured. Figure~\ref{fig:Overview}(c) shows a linewidth measurement of a resonance of a large-diameter toroid in air. Laser scanning is calibrated with a Mach-Zehnder interferometer (MZI). A linewidth measurement of the cavity in water is shown in Fig.~\ref{fig:Overview}(e). Typically the linewidth broadens by around a factor of two when the cavity is immersed in a liquid due to less mode confinement or particle binding due to impurities in the liquid.

The cavity is characterized by first positioning a large-diameter toroid resonator in a sample chamber built by gluing a glass coverslip on top of a custom made sample holder [Fig.~\ref{fig:Overview}(b)]. Water which is de-gassed in a vacuum chamber is then injected into the chamber with a syringe pump. When the chamber is filled with liquid, a tapered fiber is coupled to the resonator to inject light into the cavity. A tunable laser (New Focus TLB-6712-P) scans wavelengths from 765 nm to 781 nm while its scan wavelength is precisely calibrated by a MZI with a FSR of $17.5~\text{MHz}$. The calibration data and the transmission signal are both received by photodetectors (New Focus 1801) and monitored by a high sampling rate data acquisition card (NI PCI-6115)~\cite{yi_soliton_2015,fujii_dispersion_2020}. An example spectrum is shown in Fig.~\ref{fig:Overview}(d). More than 10 modes are excited as shown in the spectrum of approximately two FSRs. In order to introduce AMXs, we excited not only the low order modes but also other higher order modes in the cavity by placing the tapered fiber in contact with the top of the toroid~\cite{lin_wide-range_2014} and adjusting phase-matching conditions~\cite{spillane_ideality_2003}. In order to enhance the stability of the coupling condition, a wall is fabricated next to the toroid and used to support the fiber~\cite{monifi_encapsulation_2013}. The laser is tuned to a resonance by decreasing optical frequency to achieve thermal locking~\cite{carmon_dynamical_2004}. The spectrum is recorded using an optical spectrum analyzer (OSA, Thorlabs 202C).

\section{Frequency Comb Generation in Water and Air}

\begin{figure*}[ht!]
\centering
\includegraphics[width=\linewidth]{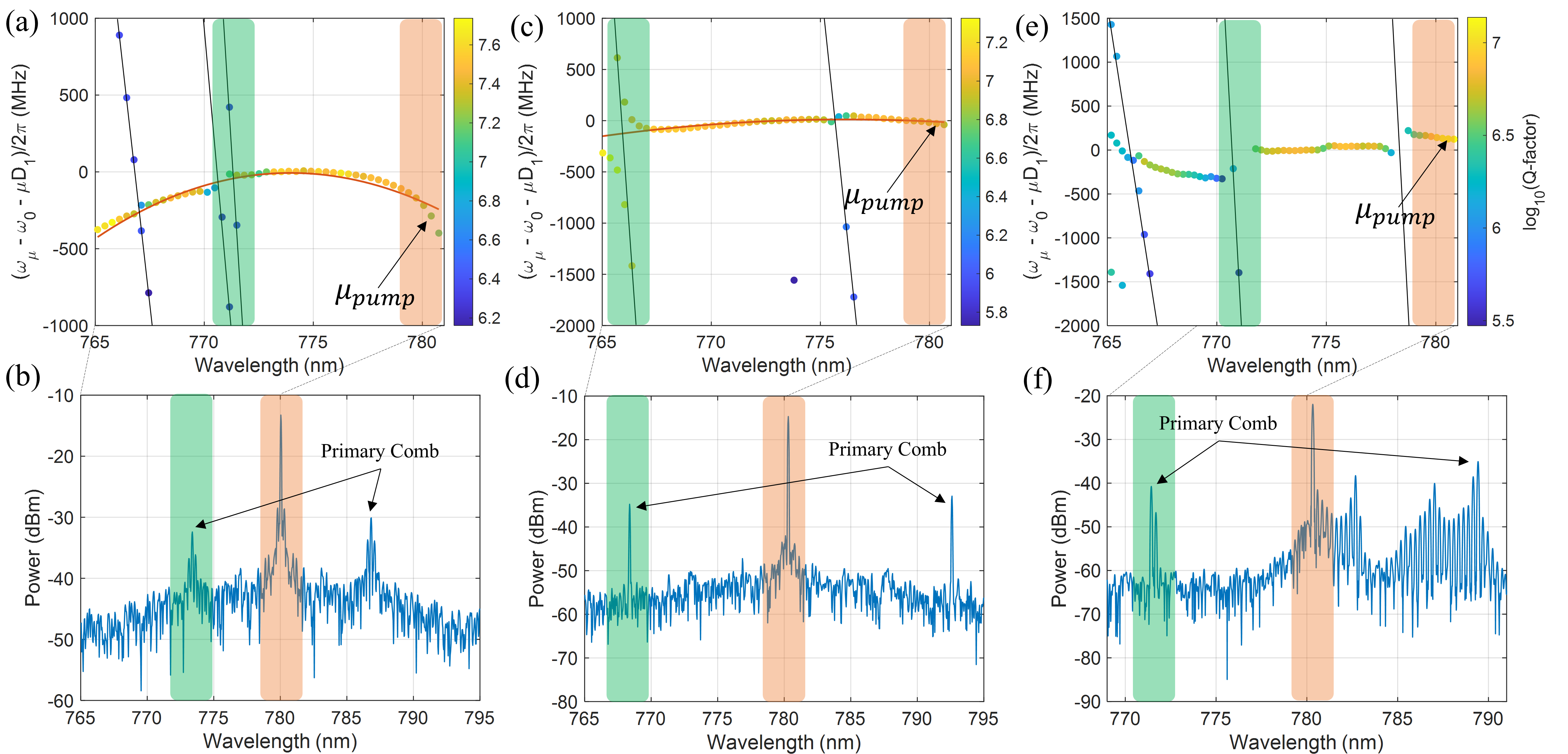}
\caption{Dispersion measurement and frequency comb generation in water. (a) Deviation of the resonance frequencies, $\omega_\mu=\omega_0+D_1\mu+\frac{1}{2}D_2\mu^2+\ldots$, from an equidistant frequency grid ($\omega_0+D_1\mu$) and $\mu$ is the relative mode number, where $D_1=2\pi\times FSR$ with respect to a pump mode ($\mu_0$). Each dot on the plot related to an eigenfrequency ($\omega_\mu$) of the cavity. A particular mode family is represented as a line that consists of colored dots corresponding to measured Q-factors on a logarithmic scale. The dot color may be used to identify a particular mode family because a mode family has similar Q. The integrated dispersion, $D_{int}=\omega_\mu-\omega_0-D_1\mu=\frac{1}{2}D_2\mu^2+\ldots$, describes normal dispersion with $D_2/2\pi=-1.1877~\text{MHz}$ (red solid line; higher-order terms are ignored). Black solid lines are drawn to visualize higher order mode families and AMXs. The AMX can be easily observed because it significantly alters eigenfrequencies, degrades the Q-factor (shown by the dot color) and transmission depth of the resonances (not shown here). The pump wavelength and AMX location is highlighted in orange and green, respectively. (b) Generated frequency comb in water when the mode indicated in (a) is pumped. The primary comb is located at the wavelength where the AMX happens (highlighted in green).(c) The integrated dispersion for a mode family with a $D_2/2\pi\sim-0.5158~\text{MHz}$. The pump wavelength and AMX location is highlighted in orange and green, respectively. (d) Generated frequency comb in water by pumping the mode shown in (c). The primary comb is located at the wavelength where the AMX happens (highlighted in green). (e) The integrated dispersion for a mode family where AMXs are considerably strong and dispersion cannot be measured precisely. The pump wavelength and AMX location is highlighted in orange and green, respectively. (f) The generated frequency comb when the mode indicated in (e) is pumped.}
\label{fig:CombWater}
\end{figure*}

Figure~\ref{fig:CombWater} shows frequency comb generation in water using a microtoroid resonator via the AMX approach. The dispersion of the cavity is characterized and plotted in Figs.~\ref{fig:CombWater}(a), (c), and (e). The measured resonance frequencies are marked as dots over the whole scan spectrum. The resonance frequencies are plotted as deviations from an equidistant frequency grid with a FSR ($D_1$). A mode family can be interpreted as a line connecting dots. If a mode family has a FSR of $D_1$ and no dispersion, it may be shown as a horizontal line. Different slopes of each line can be understood as different FSRs for each mode family. The color of the dots represent measured Q-factors and helps trace a mode family.

To better understand the eigenfrequency locations, we analyze the dispersion properties of our comb. The dispersion properties of a mode family with mode frequencies, $\omega_\mu$, can be Taylor expanded as, $\omega_\mu=\omega_0+D_1\mu+\frac{1}{2}D_2\mu^2+\ldots$, where $\mu$ is the relative mode number with respect to the pump ($\mu_0$), $\omega_\mu$ are the resonance frequencies, $\omega_0$ is the pump frequency, $D_1/2\pi$ is the FSR, and $D_2/2\pi$ is the second order dispersion (with higher order dispersion terms ignored)~\cite{herr_universal_2012}.  It is often useful to introduce an integrated dispersion, $D_{int}=\omega_\mu-\omega_0-D_1\mu=\frac{1}{2}D_2\mu^2+\ldots$, which shows the deviation of the resonance frequencies from the equidistant frequency grid (FSR = $D_1/2\pi$) with respect to a pump mode ($\mu=0$). The integrated dispersion is plotted to extract $D_2$ by fitting a curve ($D_{i>2}$ are ignored). Note that $\mu_0\neq\mu_{pump}$ for plotting purposes. The fitted dispersion coefficients are $D_1/2\pi=170.7088~\text{GHz}$, $D_2/2\pi=-1.1877~\text{MHz}$ for a family~[Fig.~\ref{fig:CombWater}(a)] and $D_1/2\pi=170.7073~\text{GHz}$, $D_2/2\pi=-0.5158~\text{MHz}$ for another mode family~[Fig.~\ref{fig:CombWater}(c)]. Dispersion for a mode family shown in Fig. 2(e) may not be estimated where a dispersion curve is distorted due to several strong AMXs present for the mode family. Because of the limited wavelength scan range and the presence of strong and weak modal couplings, an accurate measurement of these fits can not be faithfully guaranteed. The discrepancy between the fitted $D_2/2\pi\approx-1.19~\text{MHz}$ and the simulated $D_2/2\pi\approx-1.63~\text{MHz}$ can be attributed to fabrication uncertainty and the aforementioned limits.

Frequency comb generation in water was observed by pumping a resonance at $\sim780~\text{nm}$ at a pump power of $\sim20~\text{mW}$~[Fig.~\ref{fig:CombWater}(a)]. The primary comb (the first generated sideband) is located where the AMX happens~[Fig.~\ref{fig:CombWater}(b)]. A small deviation between the AMX location ($\lambda>772~\text{nm}$) and the primary comb location ($\lambda>773~\text{nm}$) can be explained by a frequency shift resulting from a temperature increase of the cavity as a function of input power and detuning. As the resonance is swept from high to low frequencies, an increase in intracavity power makes the cavity hot and the frequency shift. This alters AMX locations because different transverse modes might have different temperature shift  coefficients~\cite{miller_tunable_2015,kim_turn-key_2019}. Another mode family with a strong AMX at $\sim767~\text{nm}$ and a weak perturbation at $\sim776~\text{nm}$ is pumped~[Fig.~\ref{fig:CombWater}(c)]. As shown in~Fig.~\ref{fig:CombWater}(d), a stronger modal coupling may satisfy the phase-matching condition first over others when multiple AMXs exist even though the AMX is positioned further away from the others. Thus, the primary comb line is located at $\sim767~\text{nm}$ rather than $\sim776~\text{nm}$. Finally, the other mode family~[Fig.~\ref{fig:CombWater}(e)] is pumped where strong AMXs are present and undisturbed dispersion may not be characterized faithfully. The generated frequency comb is shown in~[Fig.~\ref{fig:CombWater}(f)] and a significant asymmetry can be observed.

\begin{figure*}[ht!]
\centering
\includegraphics[width=\linewidth]{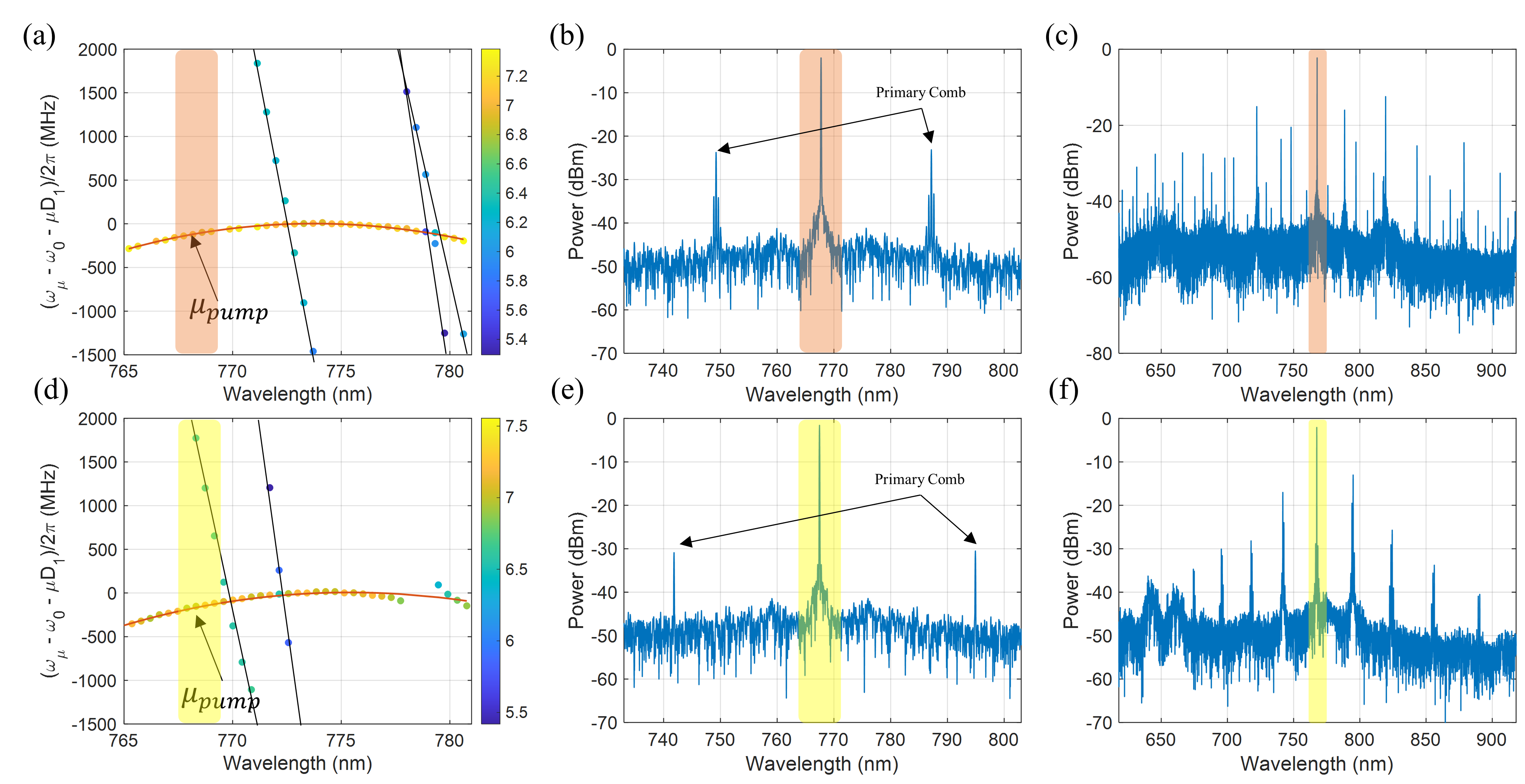}
\caption{Measured dispersion and generated frequency comb in air. (a) Integrated dispersion of a mode family where only a weak modal coupling is present in the scan range. The pump wavelength is highlighted in orange. The dot color  represents the quality factor in a logarithmic scale and helps to trace a mode family as shown in (a) and (d). (b) Primary comb lines appear beyond the wavelength scan range from the blue-detuned side. When multiple AMXs exist for the mode family, the primary comb does not appear at the closest AMX location but rather might be dependent on coupling power between the modes. (c) A broadband frequency comb generated as the laser scans from short to long wavelengths. The spectrum spans over $400~\text{nm}$ and covers the visible wavelength range.  (d) Integrated dispersion for another mode family. Several AMXs are observed over the wavelength scan range. The pump wavelength is highlighted in yellow. (e) Primary comb lines occur at $\sim30~\text{nm}$ away from the pump wavelength. (f) A broadband frequency comb spanning more than $200~\text{nm}$.}
\label{fig:CombAir}
\end{figure*}

Some applications such as environmental monitoring require sensing and identification of particles in air \cite{su_portable_2018}. Thus, a broadband frequency comb in air via an AMX is demonstrated and shown in Fig.~\ref{fig:CombAir}. Two different mode families are investigated. A high order mode family is pumped and the integrated dispersion is shown in Fig.~\ref{fig:CombAir}(a) with $D_1/2\pi$ of $214.8031~\text{GHz}$ and $D_2/2\pi$ of $-1.4099~\text{MHz}$ approximately. There is no interaction between the pump mode family and a higher-order mode family at $\lambda\sim773~\text{nm}$ where no apparent frequency shift and line broadening are observed. A weak perturbation is present at $\lambda\sim779~\text{nm}$, but primary comb lines appear at $\lambda\sim788~\text{nm}$ (48 FSRs away) which is beyond the scan range of our tunable laser~[Fig.~\ref{fig:CombAir}(b)]. It is interesting to note that when several AMXs exist at the pump wavelength, they compete to determine which is dominant in forming the comb. A strong AMX outside the scan range may explain the position of the primary comb line. As the laser is tuned from high to low frequency, a broader frequency comb is generated covering more than $400~\text{nm}$~[Fig.~\ref{fig:CombAir}(c)]. A strong AMX might also explain the broad spectrum of the frequency combs~\cite{jang_dynamics_2016}. Figure~\ref{fig:CombAir}(d) shows intergrated dispersion for a low order mode family. Fitting a curve yields $D_1/2\pi\approx214.7849~\text{GHz}$ and $D_2/2\pi\approx-1.5074~\text{MHz}$. There is a weak AMX at $\lambda\sim772~\text{nm}$ and a strong AMX $\lambda\sim778~\text{nm}$. However, it is unexpected the primary comb is not positioned at the strong AMX but rather $\sim30~\text{nm}$ (63 FSRs) away from the pump wavelength. We believe this observation can be attributed to detuning dependent AMXs and the competition between AMXs. In practice, a smaller FSR or dual comb technique would be desired to record sharp absorption features in a gas phase sample.

\begin{figure}[t!]
\centering
\includegraphics[width=.8\linewidth]{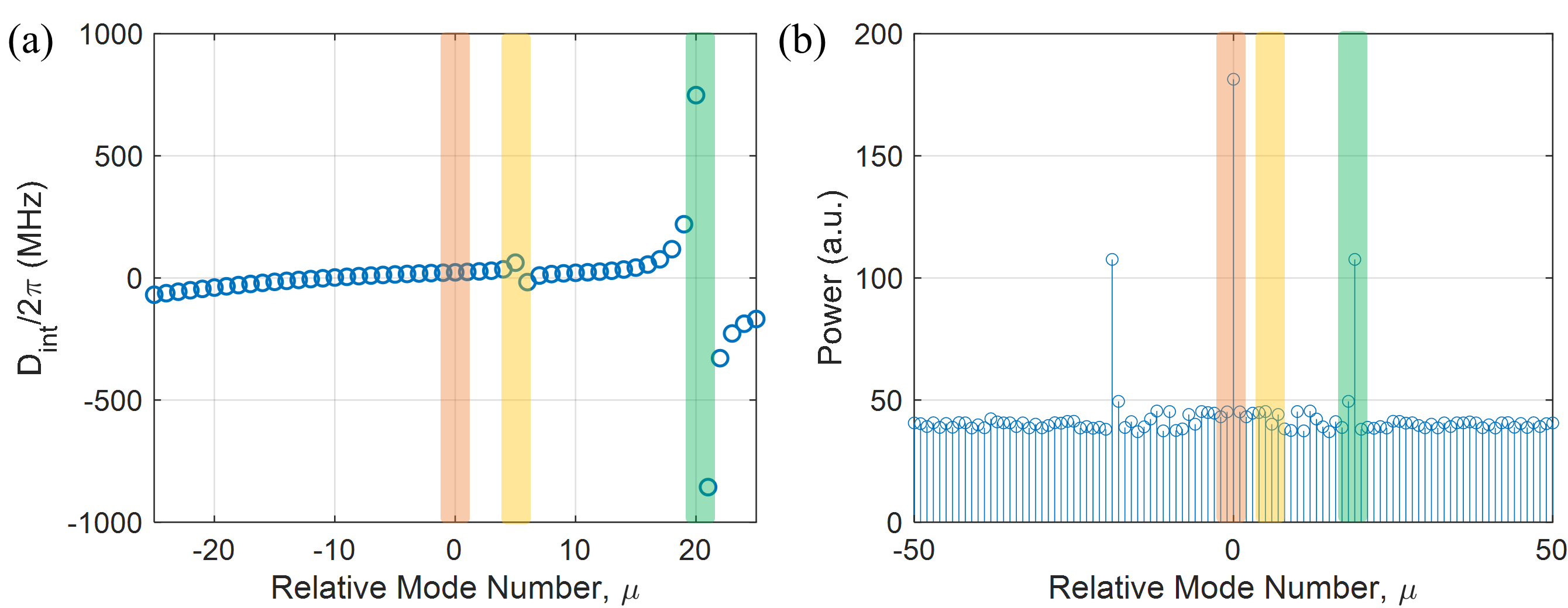}
\caption{Numerical simulation on primary comb. (a) Simulation of the integrated dispersion. Simulation parameters are described in the main text. A simple two-parameter model is used~\cite{herr_mode_2014}. The pump mode, and weak and  strong mode location are highlighted in orange, yellow, and green, respectively. (b) Primary comb line generation. The primary comb is generated where the stronger AMX happens (highlighted in green).}
\label{fig:CombSimul}
\end{figure}

In order to better understand primary comb generation in the presence of multiple AMXs for a particular pump wavelength, a numerical simulation was performed using the Lugiato-Lefever equation ~\cite{herr_temporal_2014,chembo_spatiotemporal_2013,hu_reconfigurable_2020} (see Supplementary Information Section 2 for more details). The integrated dispersion is determined using a simplified two-parameter model~\cite{herr_mode_2014} from experimental data [Fig.~\ref{fig:CombWater}(e)] and is shown in Fig.~\ref{fig:CombSimul}(a). A strong AMX is present at $\mu=20$ and a weak AMX exists at $\mu=5$. A resonance is pumped at $780~\text{nm}$ with $50~\text{mW}$. Figure~\ref{fig:CombSimul}(b) shows the primary comb lines which agree with the strong AMX location ($\mu=20$). It was found that adjusting modal coupling strengths results in different comb generation dynamics (see Supplementary Information Section 4 for more details). For example, from the simulation, by increasing the coupling strength for the weak perturbation, the primary comb lines may appear at both weak and strong AMX locations.

\section{Conclusion}

In conclusion, we generated optical frequency combs in water with a microtoroid resonator using an AMX approach. Although this demonstration is limited by naturally occurring AMXs and a few comb lines observed, it suggests that one optical resonator may function as both a spectrometer and a biosensor without building additional structures. By introducing a locking mechanism and improving the mechanical and thermal stability of the device in liquid, we believe a broadband and low phase noise frequency comb (or dark soliton) can be generated with a proper choice of pumping power and detuning~\cite{xue_mode-locked_2015}. A further investigation on AMXs was performed in air when there were multiple AMXs for a pump's mode family. It was found that if multiple AMXs are present, they compete with each other and the primary comb line location may be determined by the intermodal coupling strength. A broadband optical frequency comb spanning over $400~\text{nm}$ was generated with the aid of a strong AMX far away from the pump wavelength in the visible to NIR regime.This study suggests a path for a new multi-functional photonic device that may detect and identify a single molecule of interest in both gas and liquid.

\begin{backmatter}

\bmsection{Funding} National Institutes of Health (R35GM137988); National Science Foundation (NSF) (1842045)

\bmsection{Acknowledgements} We thank Dalziel J. Wilson for helpful reading of the manuscript.

\smallskip

\bmsection{Disclosures} J.S.: Femtorays Technologies (I)

\bmsection{Data Availability Statement} Data underlying the results presented in this paper are not publicly available at this time but may be obtained from the authors upon reasonable request.

\bmsection{Supplemental document}See Supplement 1 for supporting content.
\end{backmatter}

\bibliography{MyLibrary}

\end{document}


\maketitle

\section{Dispersion engineering on a microbubble resonator}

\begin{figure}[b!]
\centering
\includegraphics[width=\linewidth]{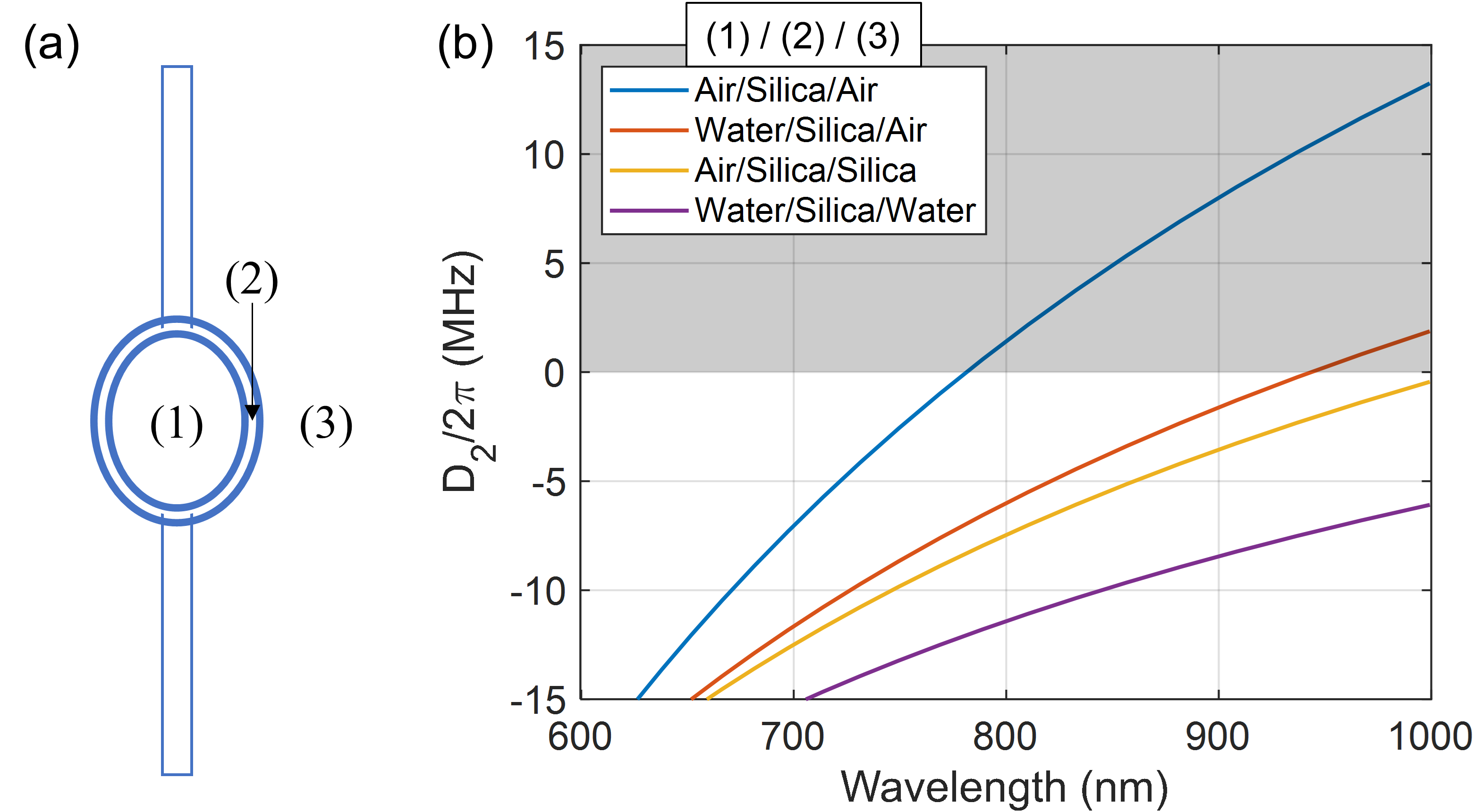}
\caption{Finite element simulation on a microbubble resonator. (a) Geometry of a microbubble resonator. Inside (1) and outside (3) of the resonator can be air or liquid. Silica is chosen for material of the resonator (2). (b) Simulated second order dispersion parameters at various wavelengths. A diameter of a microbubble resonator is 120 µm and thickness of a wall is 1.5 µm in the simulation.}
\label{fig:Bubble}
\end{figure}

It is demonstrated that engineering geometry of a microbubble resonator can yield anomalous dispersion at near-IR wavelength~($\sim780~\text{nm}$)~\cite{yang_four-wave_2016}. As a thickness of a microbubble resonator gets thinner, the light gets confined tighter in the waveguide. However, this demonstration is performed in air and injecting liquid into or over the resonator can degrade the confinement, resulting in normal dispersion at near-IR wavelength~($\sim780~\text{nm}$). Figure~\ref{fig:Bubble} shows second order dispersion parameters of a silica microbubble resonator with a diameter of 120 µm and thickness of 1.5 µm for given boundary conditions. It can be clearly observed that as the surrounding material is replaced by liquid from air anomalous dispersion~($D_2>0$) at$\sim780~\text{nm}$ cannot be achieved.

\begin{figure}[b!]
\centering
\includegraphics[width=\linewidth]{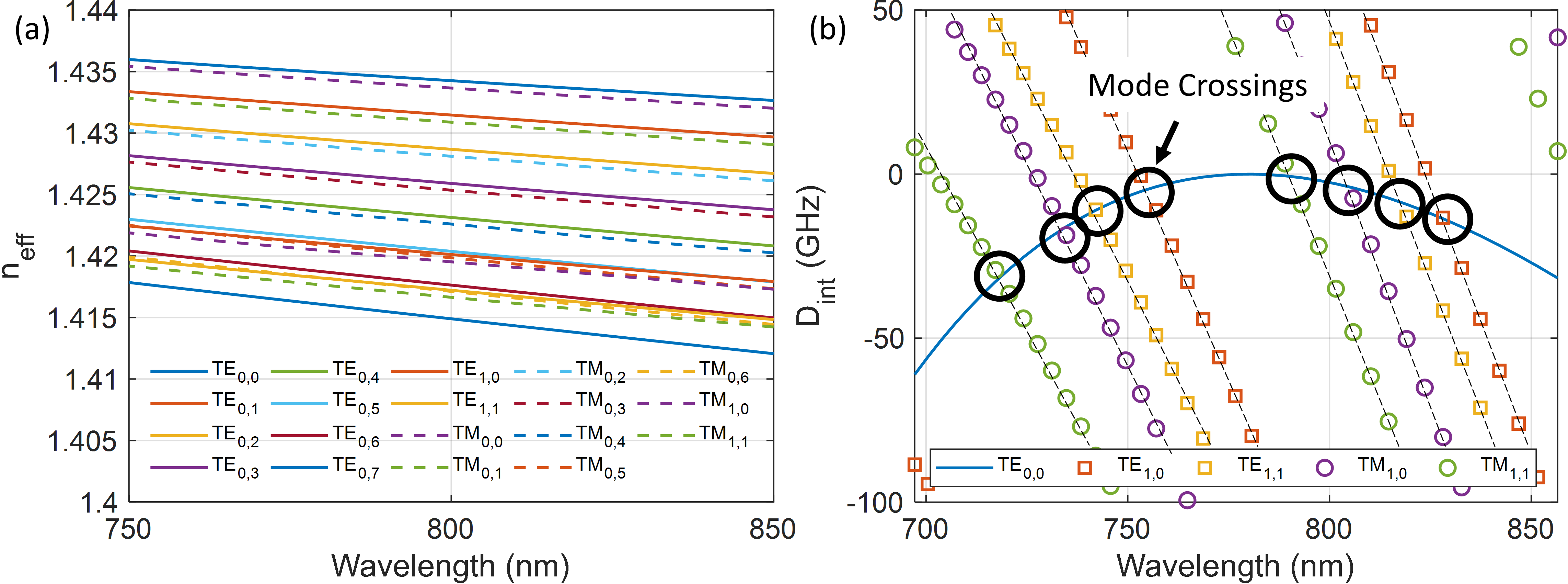}
\caption{(a) Mode families of a microtoroid resonator. Effective refractive indices of a toroid resonator with a major diameter of $300~\mu\text{m}$ and minor diameter of $30~\mu\text{m}$. (b) Integrated dispersion with respect to the fundamental TE mode. Mode crossings with higher-order optical modes are observed.}
\label{fig:Modal}
\end{figure}

\section{Lugiato-Lefever Equation (LLE) Simulation}

The master equation can be written as ~\cite{herr_temporal_2014,guo_universal_2017}:
\begin{equation}
\frac{\partial A}{\partial t}-i\frac{1}{2}D_2\frac{\partial^2A}{\partial\phi^2}-ig|A|^2A=-\left(\frac{\kappa}{2}+i(\omega_0-\omega_p)\right)A+\sqrt{\frac{\kappa\eta P_{in}}{\hbar\omega_0},}
\label{LLE}
\end{equation}
where $A(\phi,t)$ is the internal electric field within the resonator. $\phi$ is the azimuthal coordinate around the resonator. $\kappa=\kappa_{ex}+\kappa_0$ is the total cavity loss rate, where $\kappa_{ex}$ and $\kappa_0$ are the coupling rate and intrinsic loss rate. $\eta=\kappa_{ex}/\kappa$ is the coupling efficiency. $\omega_0$ and $\omega_p$ are the resonance frequency and the pumping frequency. $g=\hbar\omega _0^2cn_2/n^2V_0$ is the Kerr frequency shift per photon, where $n$ is the refractive index, $n_2$ is the nonlinear optical index, and $V_0$ is the effective mode volume. $D_2$ is the second order dispersion parameter. As discussed in the main text, we ignore higher-order ($D_{i>2}$) dispersion parameters to simplify simulations.

Simulation parameters are obtained from a dispersion measurement and a finite element simulation: $D_1/2\pi=200~\text{GHz}$, $\kappa=\kappa_{ex}/2=\kappa_0/2=24~\text{MHz}$, $V_0=10866~\mu\text{m}^3$, $P_{in}=50~\text{mW}$, and $\lambda_0=780~\text{nm}$. A two-parameter model is used to simulate avoided mode crossings (AMXs) as introduced in~\cite{herr_mode_2014}. The two parameters, $a$ and $b$, illustrate the coupling strength and an AMX location, respectively. As shown in Fig.~4(a), we introduced one weak and one strong modal crossing in the integrated dispersion ($D_{int}$) as experimentally measured (similar to what is shown in Fig.~2(c) but not exactly the same). The model parameters $a$ is $\kappa/2$, $10\kappa$, and $b$ is $5$, $20$ for the weak and strong AMXs, respectively. The resonance is swept from the blue-detuned side to the red-tuned side. The simulation is stopped when a primary comb is observed as shown in Fig.~4(b).

\section{Mode Interactions}

A finite element simulation is performed to find the effective refractive indices of 19 different optical modes in a toroid resonator with a major diameter of $300~\mu\text{m}$ and minor diameter of $30~\mu\text{m}$ as a function of wavelength~[Fig.~\ref{fig:Modal}(a)]. The integrated dispersion which reveals second-order or higher dispersion parameters is plotted based on the simulation results~\cite{herr_mode_2014} and mode crossings can be observed~[Fig.~\ref{fig:Modal}(b)]. 

\section{Coupling Strength and Broadband Frequency Combs}

\begin{figure*}[ht!]
\centering
\includegraphics[width=\linewidth]{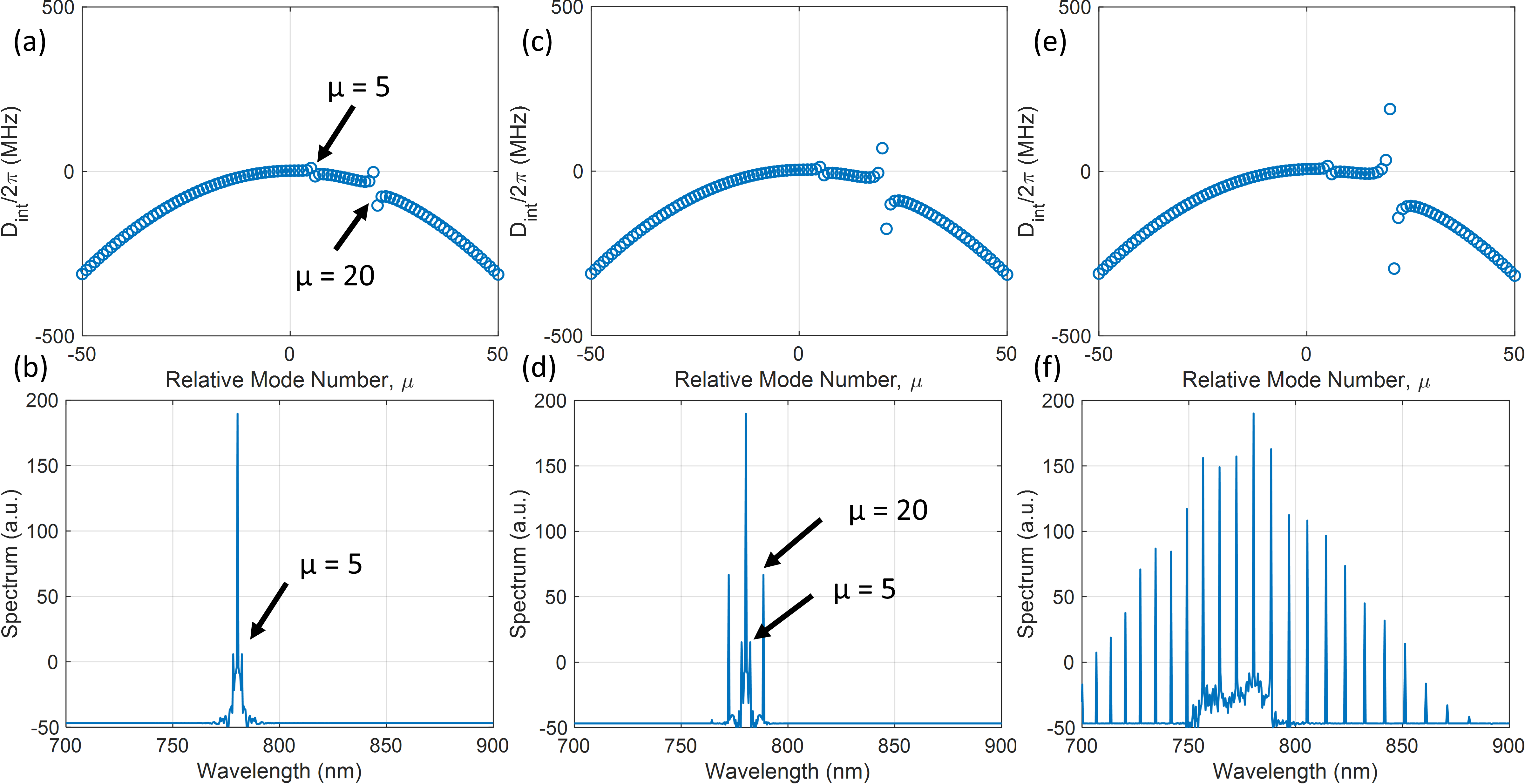}
\caption{(a, c, e) Integrated dispersion with a weak perturbation at $\mu=5$ and a strong modal crossing at $\mu=20$. Coupling strength increased from left to right for the strong modal crossing. (b, d, f) Simulated spectrum based on the integrated dispersion input from (a, c, e), respectively. A competition for the four-wave mixing process between weak and a strong modal couplings can be observed in (b) and (d). A broadband frequency comb is generated when the coupling strength is further increased as shown in (f).}
\label{fig:Sim}
\end{figure*}

Since a whispering gallery mode toroid resonator can support more than ten optical modes, it may by expected that multiple AMXs can be present with respect to a pumping mode as shown in Fig.~\ref{fig:Modal}. Based on the LLE simulation, we investigated a competition for the four-wave mixing (FWM) process and the relationship between the intermodal coupling strength and the bandwidth of a frequency comb. Figure~\ref{fig:Sim}(a) shows the integrated dispersion with a weak modal crossing ($a=\kappa/2$ and $b=5$) and a stronger AMX ($a=2\kappa$ and $b=5$). Other simulation parameters are written in section 2 and fixed for all the simulations shown here. As shown in Fig.~\ref{fig:Sim}(b), the weak modal crossing initiated sideband generation first and no sideband generation at the strong AMX location is observed. This may be understood as when an AMX location is far away from a pump, it requires more frequency shift to meet the phase-matching condition for the FWM process. Next, the coupling strength for the strong AMX increased by $a=5\kappa$~[Fig.~\ref{fig:Sim}(c)]. Then, the strong AMX met the phase-matching requirement and a comb was generated at both AMX locations. Finally, we increase the coupling strength for the strong AMX by $a=10\kappa$~[Fig.~\ref{fig:Sim}(e)]. Stronger modal coupling generated a broader frequency comb as shown in Fig~\ref{fig:Sim}(f)~\cite{jang_dynamics_2016}.

\bibliography{MyLibrary}